# High-efficiency terahertz spin-decoupled meta-coupler for spoof surface plasmon excitation and beam steering

Li-Zheng Yin, Tie-Jun Huang, Feng-Yuan Han, Jiang-Yu Liu, and Pu-Kun Liu

**ABSTRACT** Spoof surface plasmon (SSP) meta-couplers that efficiently integrate other diversified functionalities into a single ultrathin device are highly desirable in the modern microwave and terahertz fields. However, the diversified functionalities, to the best of our knowledge, have not been applied to circular polarization meta-couplers because of the spin coupling between the orthogonal incident waves. In this paper, we propose and numerically demonstrate a terahertz spin-decoupled bifunctional meta-coupler for SSP excitation and beam steering. The designed meta-coupler is composed of a coupling metasurface and a propagating metasurface. The former aims at realizing anomalous reflection or converting the incident waves into SSP under the illumination of the left or right circular polarization waves, respectively, and the latter are used to guide out the excited SSP. The respective converting efficiency can reach 82% and 70% at 0.3THz for the right and left circular polarization incident waves. Besides, by appropriately adjusting the reflection phase distribution, many other diversified functionalities can also be integrated into the meta-coupler. Our study may open up new routes for polarization-related SSP couplers, detectors, and other practical terahertz devices.

## I. INTRODUCTION

Metasurfaces are 2D metamaterials capable of manipulating electromagnetic (EM) waves at will by imparting space-variant phase changes on the incident waves. Owing to the unique characteristics of subwavelength thickness and strong ability in wavefront control, metasurfaces have shown many potential applications in beam focusing [1-6], multichannel reflection [7-10], waveplates [11-15], holograms [16-19], vortex generation [20], and polarimeters [21-23]. In particular, by adjusting the phase gradient appropriately, metasurfaces can act as bridges linking the traveling waves and the surface waves [24]. In addition, by periodically arranging the unit cells, metasurfaces can be transformed to spoof surface plasmon (SSP) waveguides [25]. The corresponding SSP modes and dispersion characteristics can be tuned by elaborately modulating the propagating unit cells. This property increases the flexibility in designing SSP waveguides. Moreover, compared with other SSP couplers, metasurfaces also have advantages in miniaturization because the thickness is far smaller than the wavelength in free space. Various SSP meta-couplers (couplers based on metasurfaces) have been designed and fabricated in the past few years [24-28].

Recently, polarization-controlled bifunctional SSP and surface plasmon (SP) meta-couplers are proposed and experimentally demonstrated in the microwave and optical regimes [29-31]. These works open up new routes for multi-freedom-controlled SSP/SP excitation. However, most of them are designed with symmetrical or similar functionalities, not reaching the goal of realizing independent multiple functionalities. SSP/SP couplers that integrate multiple diversified functionalities into a single ultrathin device are highly desirable in the modern microwave, terahertz, and optical fields. In 2017, by using the unit cells with different reflection phase, Cai *et.al.* design high-performance reflective bifunctional metasurfaces with independent beam steering and SSP excitation for orthogonal linear-polarization incidence in the microwave regime [32]. Similar functionalities are also realized for transmissive metasurfaces with independent anomalous reflection and beam focusing. In addition, at visible wavelength, Ding *et.al.* use gap-plasmon metasurfaces to implement polarization-controlled unidirectional SP excitation and beam steering at normal incidence [33]. Similarly, polarization-switchable multi-functional metasurface for focusing and SP excitation are also designed by Ling *et.al.* [34]. All these works realize the independent SSP/SP excitation and beam steering under orthogonal linear polarization electromagnetic wave (EMW) illumination. However, meta-couplers that integrate multiple diversified functionalities, to the best of our knowledge, have not been demonstrated for circular polarization (CP) incidence up to now because of spin coupling between the orthogonal incident waves [35, 36].

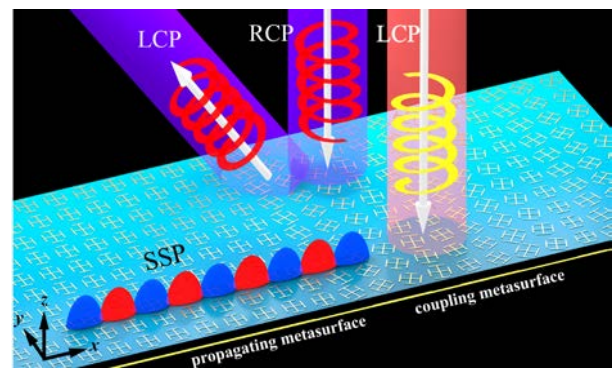

Figure 1. Schematic view of the bifunctional meta-coupler that can realize CP-controlled unidirectional SSP excitation and anomalous reflection. The meta-coupler consists of a coupling metasurface and a propagating metasurface.

Table 1. Exact structure parameters of the unit cells.   unit: μm

| Unit cell number | $b_x$ | $l_y$ | $b_y$ | $l_x$ |
|---|---|---|---|---|
| 1 | 120 | 163 | 103.3 | 140 |
| 2 | 110 | 145 | 101.7 | 135 |
| 3 | 106.7 | 141.7 | 86.7 | 130 |
| 4 | 103.3 | 140 | 120 | 163 |
| 5 | 101.7 | 135 | 110 | 145 |
| 6 | 86.7 | 130 | 106.7 | 141.7 |

In this work, we design and demonstrate a high-efficiency CP-controlled bifunctional meta-coupler for independent SSP excitation and anomalous reflection. The bifunctional meta-coupler is composed of a coupling metasurface and a propagating metasurface, as illustrated by Fig. 1. By combining the resonant phase with the Pancharatnam–Berry phase, efficient decoupling can be realized between orthogonal CP incident waves [37-39]. In this way, arbitrary independent phase gradients can be achieved theoretically. By elaborately adjusting the phase distribution of the coupling metasurface, the incident waves can be converted into SSP with longitudinal wave vector $k_x = 1.2\ k_0$ or anomalously reflected with reflection angle $\theta_r = 13.9º$ under the illumination of the left circular polarization (LCP) or the right circular polarization (RCP) beam, respectively. Finite element method (FEM) full-wave simulations show that the respective converting efficiency of the RCP and LCP at 0.3 THz can reach up to 82% and 70% and their corresponding half-power bandwidth are 0.0285 and 0.01 THz. Besides, we further demonstrate that the meta-coupler can maintain high efficiencies (>48%) with the incident Gaussian beam waist-width ranging from 0.7λ to 1.7λ, which verifies the robustness of the designed bifunctional meta-coupler. In addition to the anomalous reflection, many other diversified functionalities can also be integrated into the SSP meta-coupler in this way.

## II.  THEORY ANALYSIS

We now describe the physical mechanism of spin-decoupling between orthogonal CPs. First of all, we analyze the electromagnetic properties of the unit cells of the metasurface. Modeling the reflective metasurface as a one-port device, its reflection properties can be characterized by a Jones matrice $\mathbf{R} = \begin{bmatrix} r_{xx} & r_{xy} \\ r_{yx} & r_{yy} \end{bmatrix}$. The coordinate base $\{\hat{x}, \hat{y}\}$ of the matrice $\mathbf{R}$ represents the unit vectors of the orthogonal linear polarization EMWs. If we consider the unit vectors of circular polarized waves $\hat{e}_\pm = (\hat{x} \pm i\hat{y})/\sqrt{2}$ as the new coordinate base, then the new reflective Jones matrix $\mathbf{R}(0)$ can be expressed as

$$\mathbf{R}(0) = \begin{bmatrix} \frac{1}{2}(r_{xx}+r_{yy})+\frac{i}{2}(r_{xy}-r_{yx}) & \frac{1}{2}(r_{xx}-r_{yy})-\frac{i}{2}(r_{xy}+r_{yx}) \\ \frac{1}{2}(r_{xx}-r_{yy})+\frac{i}{2}(r_{xy}+r_{yx}) & \frac{1}{2}(r_{xx}+r_{yy})-\frac{i}{2}(r_{xy}-r_{yx}) \end{bmatrix} \quad (1)$$

To simplify the complexity of the further analysis, we design the unit cells with mirror symmetry structures. This can efficiently eliminate the cross coupling between orthogonal linear-polarization incident waves, i.e. $r_{xy} = r_{yx} = 0$ [36]. In this case, the Jones matrix $\mathbf{R}(0)$ can be written as

$$\widetilde{\mathbf{R}}(0) = \begin{bmatrix} \frac{1}{2}(r_{xx}+r_{yy}) & \frac{1}{2}(r_{xx}-r_{yy}) \\ \frac{1}{2}(r_{xx}-r_{yy}) & \frac{1}{2}(r_{xx}+r_{yy}) \end{bmatrix}, \quad (2)$$

where the figure "0" in bracket represents the rotation angle of the unit cells. When we rotate the unit cells at an arbitrary angle $\theta$, a necessary rotation transformation will be made on the original matrix. Utilizing the rotation operator $\mathbf{Q}(\theta) = \begin{bmatrix} e^{j\theta} & 0 \\ 0 & e^{-j\theta} \end{bmatrix}$, we get

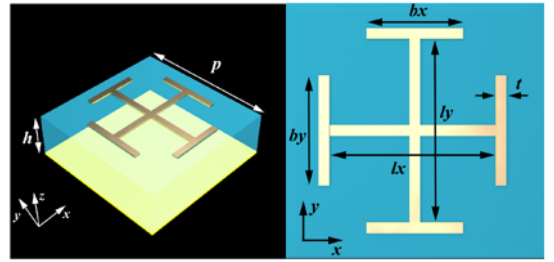

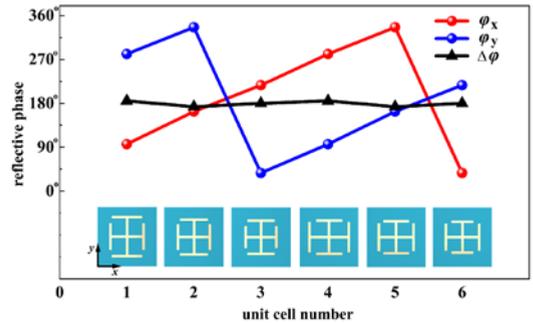

Figure 2.   (a) Schematic of the unit cell of the coupling metasurface. The period and thickness of the unit cell are 233μm and 50μm, respectively. The exact structure parameters of the unit cells used in this work are listed in Tab. 1 and their respective reflection phase under x- and y-polarized incidence are plotted in (b), where Δφ = |φx − φy| approximately equal to 180º for all unit cells to guarantee that rxx= − ryy.

$$\widetilde{\mathbf{R}}(\theta) = \mathbf{Q}(\theta)\widetilde{\mathbf{R}}(0)\mathbf{Q}^{\dagger}(\theta) = \begin{bmatrix} \frac{1}{2}(r_{xx}+r_{yy}) & \frac{1}{2}(r_{xx}-r_{yy})e^{-j2\theta} \\ \frac{1}{2}(r_{xx}-r_{yy})e^{j2\theta} & \frac{1}{2}(r_{xx}+r_{yy}) \end{bmatrix}, \quad (3)$$

where $\widetilde{\mathbf{R}}(\theta)$ represents the Jones matrice of the unit cell with a rotation angle $\theta$ under the circular polarized mode base, and $\mathbf{Q}^{\dagger}(\theta)$ represents the transposed conjugation of the rotation operator $\mathbf{Q}(\theta)$. Assuming that the reflection coefficient $r_{xx}$ and $r_{yy}$ meet the relation $r_{xx} = -r_{yy}$, then the Jones matrice $\widetilde{\mathbf{R}}(\theta)$ is simplified as

$$\widetilde{\mathbf{R}}(\theta) = \begin{bmatrix} 0 & r_{xx}e^{-j2\theta} \\ r_{xx}e^{j2\theta} & 0 \end{bmatrix}. \quad (4)$$

For the CP incident waves with opposite handness, i.e. $\begin{bmatrix} 1 \\ 0 \end{bmatrix}$ and $\begin{bmatrix} 0 \\ 1 \end{bmatrix}$, the respective reflected waves are $\begin{bmatrix} 0 \\ r_{xx}e^{j2\theta} \end{bmatrix}$ and $\begin{bmatrix} r_{xx}e^{-j2\theta} \\ 0 \end{bmatrix}$. Compared with the incident waves, we can find that, in addition to the polarization transformation, two opposite geometrical angles are encoded as phase shifts (Pancharatnam–Berry phase) into the reflective CP waves. The opposite phase shifts for the orthogonal CPs lead to symmetrical functionalities, which is also called photonics spin hall effect (PHSE) [35, 36]. To break the symmetry and realize the independent phase control for orthogonal CP incidence, another coefficient $r_{xx}$ is considered as an additional variable to control the reflected waves. Based on the previous assumption $r_{xx} = -r_{yy}$, we further consider that $r_{xx} = -r_{yy} = e^{j\varphi}$, which means that each unit cell has a specified resonant phase with uniform electric field amplitude. To meet this condition, the losses of the metal and dielectric spacer are small enough and the working frequency is beyond the resonant absorption band. In this work, the amplitude of the reflection coefficient of the unit cells in use is all higher than 0.99, as illustrated by Fig. 3(b). In this way, another resonant phase $\varphi$ is introduced as a new degree of freedom. Therefore, the Jones matrice can be expressed as

$$\widetilde{\mathbf{R}}(\theta) = \begin{bmatrix} 0 & e^{j(\varphi-2\theta)} \\ e^{j(\varphi+2\theta)} & 0 \end{bmatrix}. \quad (5)$$

Then the reflected waves under the LCP and RCP incidence can be expressed as $\begin{bmatrix} e^{j(\varphi+2\theta)} \\ 0 \end{bmatrix}$ and $\begin{bmatrix} 0 \\ e^{j(\varphi-2\theta)} \end{bmatrix}$, respectively. By elaborately adjusting the resonant phase $\varphi$ and the rotation angle $\theta$, arbitrary independent reflection phase can be obtained for orthogonal CP wave incidence.

## III. OVERALL DESIGN OF THE META-COUPLER
### A. DESIGN OF THE COUPLING METASURFACE

To realize CP-controlled SSP excitation and beam steering, 2D arrays of unit cells with different resonant phase and rotation angles are required to constitute the coupling metasurface. In ths work, $f = 0.3$ THz is chosen as the central working frequency and the period of the coupling unit cells is set as $p = 233\mu m$. Besides, we select $\Delta\varphi = 60°$ and $\Delta\theta = 20°$ so the total reflection phase difference between the adjacent coupling unit cells for the LCP and RCP incidence is 100° and 20°, respectively. Of course, the selections of $\Delta\varphi$ and $\Delta\theta$ are arbitrary as long as they meet the conditions that $|\Delta\varphi - 2\Delta\theta| > k_0 p > |\Delta\varphi + 2\Delta\theta|$ or $|\Delta\varphi + 2\Delta\theta| > k_0 p > |\Delta\varphi - 2\Delta\theta|$, where $k_0$ represents the wave vector in free space. For the RCP incidence, the longitudinal wave vector of reflected waves is $k_r = (\Delta\varphi - 2\Delta\theta)/p = 0.24\, k_0$. Therefore, the anomalous angle $\theta_r = \arcsin(k_r/k_0) = 13.9°$ under normal incidence. For the LCP incidence, the longitudinal wave vector of reflected waves is $k_r = (\Delta\varphi + 2\Delta\theta)/p = 1.2\, k_0$, which means that the surface waves are excited on the coupling metasurface. By designing a propagating metasurface with the same Eigen wave vector and electric field modes, the excited surface waves can be guided out with high efficiency. We now design the practical unit cells that meet all the theoretical requirements mentioned above. The unit cells are composed of metal patches and flat metal mirror separated by a 50 μm-thick dielectric spacer Taconic RF-43 with the relative permittivity $\varepsilon_r = 4.3 - 0.01i$. The metal patches and mirrors are composed of copper with a thickness of 2 μm. The permittivity of copper is characterized by a Drude model $\varepsilon(\omega) = 1 - \omega_p^2 / (\omega^2 + i\omega\gamma)$, where ω is the angular frequency, $\omega_p = 1.123 \times 10^{16}$ Hz is the plasma frequency and $\gamma = 1.379 \times 10^{13}$ Hz is the collision frequency [40, 41]. We adopt the conventional crossed "H" structures as the metal patches and the width of each rod is $t = 10$ μm, as illustrated by Fig. 2(a). The symmetrical structure guarantees that the cross coupling between the $x$- and $y$-polarized EMWs can be eliminated efficiently. Besides, this structure also has an advantage in that the resonant phase under the illumination of $x$- and $y$-polarized EMWs can be independently controlled. After determining the structures of the unit cells, we next obtain the relations between the resonant phase and the structure parameters of the unit cells with the help of FEM simulation. Compared with the conventional design process, there are more restricted requirements on the unit cells in this work. The patches can not be too large or it will affect the rotation and cause the mutual coupling between the adjacent unit cells. Under this condition, by sweeping the key structure parameters, we find that the unit cells can still provide a sufficiently large resonant phase range to design the coupling metasurface. We choose six unit cells as a super cell and the exact structure parameters of the unit cells are illustrated in Tab.1. The reflection amplitude and phase versus frequency under $x$-polarized EMW illumination are plotted in Figs. 3(a), and 3(b), from which we can find that the high reflection efficiencies and linear phase gradients are both realized at 0.3 THz. Figure 2(b) shows the reflection phase of each unit cell under the illumination of $x$- and $y$-polarized EMWs. $\Delta\varphi$, which approximatively equal to 180° for all unit cells, denote the difference between $\varphi_{xx}$ and $\varphi_{yy}$. This approximate constant guarantees that $r_{xx} = -r_{yy}$.

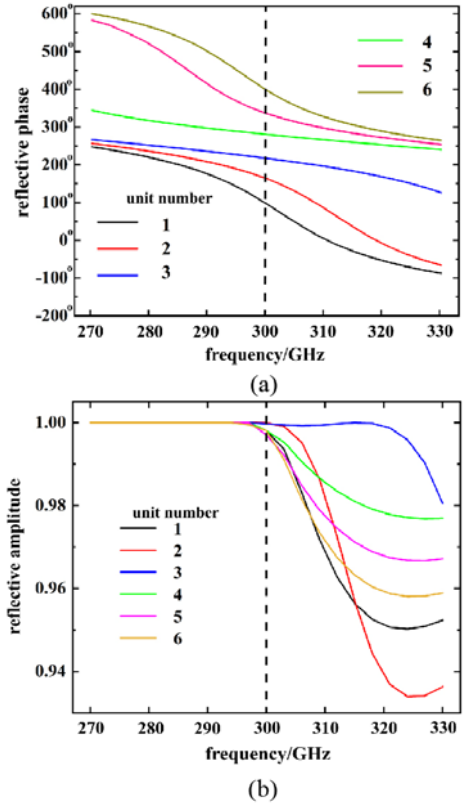

Figure 3. The reflection phase (a) and amplitude (b) of the selected unit cells versus frequency under the illumination of x-polarized EMWs.

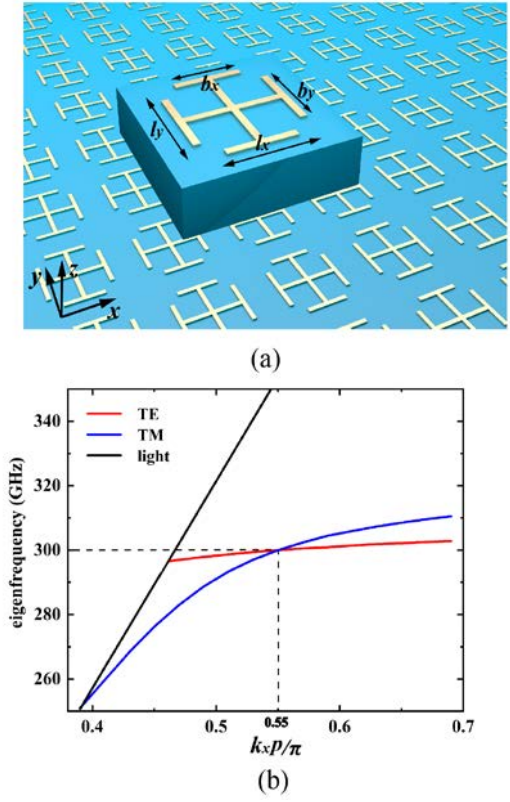

Figure 4. (a) Schematic of the unit cells of the propagating metasurface. The length of the four rods are bx = 95 μm, lx = 133 μm, by = 107 μm, and ly = 150 μm. The period and thickness of the unit cell are 233μm and 50μm, respectively. (b) The dispersion curves of the TE and TM mode SSP.

B. DESIGN OF THE PROPAGATING METASURFACE

Next, we will design the propagating metasurface so that the excited surface waves can be guided out efficiently. The propagating metasurface is periodic in the $x$- and $y$-directions with the period $p = 233$ μm. In addition to the wave vector match condition, to further improve the converting efficiency, the unit cells of the propagating metasurface should also be designed similar to those of the coupling metasurface in shape. In this way, the mode mismatch between the two parts can be efficiently weakened and a smooth and efficient transition between the excited surface waves and SSP can be realized. Based on the targets mentioned above, the unit cells which meet all the requirements are illustrated by Fig. 4(a). Similar to those of the coupling metasurface, the unit cells of the propagating metasurface are tri-layer structures consisting of flat copper mirror and patches, with the same dielectric embedded between them. Each patch is composed of two vertically crossed "H" structures and the width of each rod is $t = 10$ μm. The optimal structure parameters of unit cells are $b_x = 95$ μm, $l_x = 133$ μm, $b_y = 107$ μm, and $l_y = 150$ μm. The propagating metasurface composed of this unit cell supports both TE and TM mode SSP. The electric current and electric field distributions of the TE and TM mode SSP are plotted in Figs. 5(a), and (b), respectively. In order to meet the wave vector match condition, the dispersion curves of TE and TM modes should cross at the frequency $f = 0.3$ THz with the desired longitudinal wave vector, as illustrated by Fig. 4(b). As can be seen from the dispersion relations, the curve of TM SSP overlaps with light line in the beginning and finally deviates from it while the curve of the TE SSP origins from 0.295 THz and is relatively flat in the whole frequency band. In the same range of $k_x$, the frequency band of TM SSP is far wider than that of the TE SSP.

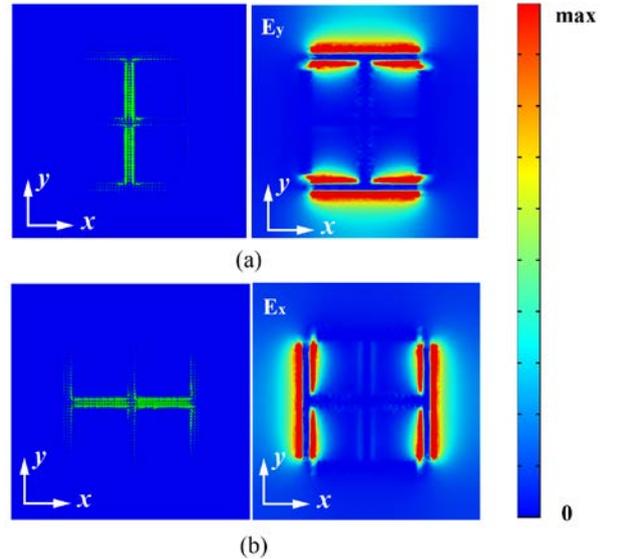

Figure 5. The electric current and electric field distributions of TE (a) and TM (b) mode SSP.

## IV. RESULTS

### A. SIMULATION RESULTS OF UNIDIRECTIONAL SSP EXCITATION AND ANOMALOUS REFLECTION

With both the coupling metasurface and propagating metasurface designed, finally, we demonstrate the functionalities of CP-controlled SSP excitation and anomalous reflection. Figure 6(a) is a schematic view of the complete structure of the meta-coupler. A Gaussian beam with waist-width $w = 1.2 \lambda$ is used as the incident plane wave source. For the RCP incidence, the incident waves "see" a small phase gradient ($k_x = 0.24\ k_0$) and are anomalously reflected with the reflection angle $\theta_r = 13.9°$. The scattering electric field components in the $x$- and $y$-directions are illustrated by Figs. 6(b), and (c), which demonstrate the function of anomalous reflection. Comparing $\mathbf{E}_x$ with $\mathbf{E}_y$, we can also find that the chirality of the incident waves changes.

For the LCP incidence, a large phase gradient ($k_x = 1.2\ k_0$) will be added on the incident waves, so the surface waves will be excited on the coupling metasurface. Due to the matches in longitudinal wave vector and electric field modes, the excited surface waves can be efficiently guided out from the coupling metasurface to the propagating metasurface, as illustrated by Figs. 6(d), and (e). The visual electric field distributions are in agreement with theory, confirming the function of efficient SSP excitation. Next, we will quantitatively demonstrate the working performance of the designed meta-coupler.

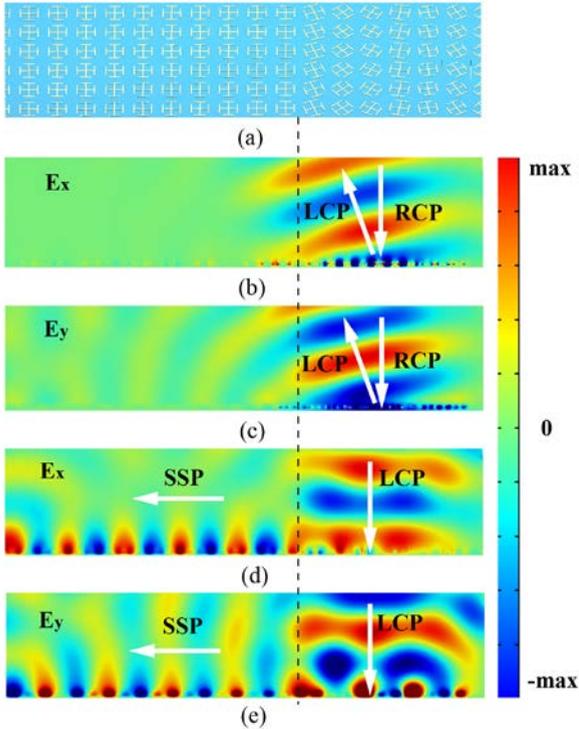

Figure 6. (a) Schematic view of the complete structure of the designed bifunctional meta-coupler. (b) and (c) are respective scattering electric field components in the x- and y-directions under the illumination of RCP waves. The RCP incident waves are transformed into LCP waves and anomalously reflected with the reflection angle θr = 13.9°. (d) and (e) are electric field components in the x- and y-directions under the illumination of the LCP waves. Surface waves are excited on the coupling metasurface and guided out from the propagating metasurface. The black dotted line represents the boundary of the coupling metasurface and propagating metasurface regions.

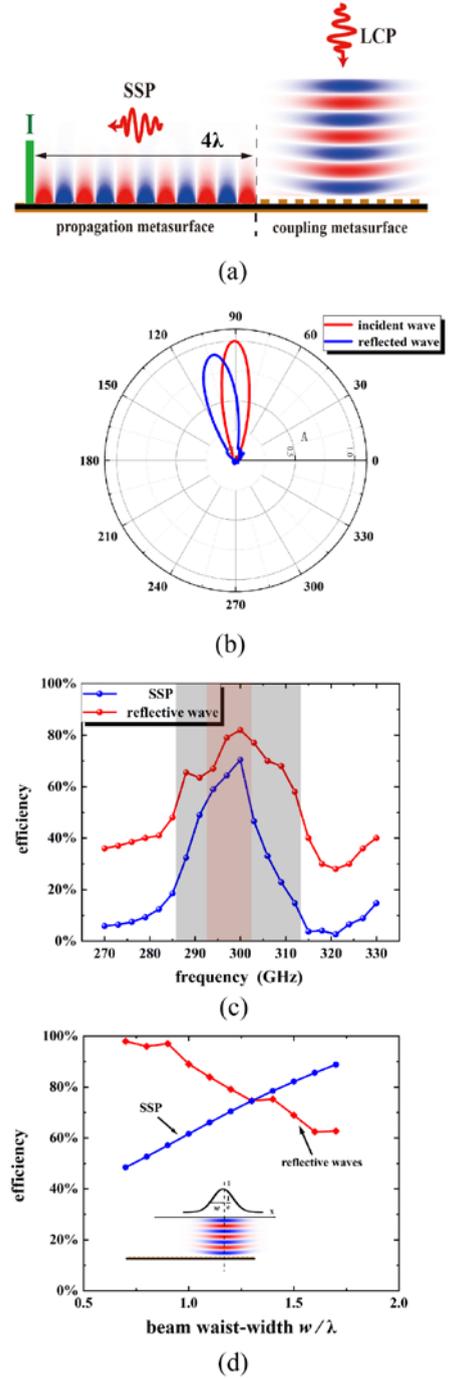

Figure 7. (a) Schematic of SSP excitation under the LCP illumination, where "I" represents the integral position of power of SSP. (b) The far-field patterns of the incident and reflected waves for the RCP incidence. (c) The converting efficiencies of the LCP and RCP incidence under the condition that the waist-width w = 1.2 λ, which are plotted by blue and red curves, respectively. The different shadow colors represent the corresponding frequency intervals that the excitation efficiencies are over 50%. (d) The relations between the converting efficiencies and incident Gaussian beam waist-width w. The inset illustrates the definition of waist-width. In the range from 0.7 λ to 1.7λ, the bifunctional metasurface can maintain high efficiency (>48%) for both the LCP and RCP incidence.

B. CONVERTING EFFICIENCIES OF THE BIFUNCTIONAL META-COUPLER

The most important indicator to evaluate the performance of the meta-coupler is the converting efficiency. In this work, due to the difference of the scattering electric field distributions of the polarization-controlled functionalities, the converting efficiencies for the LCP and RCP incidence are calculated in different methods. For the RCP incidence, the reflected waves in the far field and the incident waves are both Gaussian traveling waves. Therefore, the square of the ratio of reflected waves to incident waves, i.e. $|\mathbf{E}_r|^2 / |\mathbf{E}_i|^2$ ($\mathbf{E}_r$ and $\mathbf{E}_i$ are the respective electric field amplitude of the reflected and incident Gaussian waves), can be used to evaluate the efficiency [33]. The far-field patterns of the incident and reflected waves at 0.3 THz are plotted in Fig. 7(b). In this way, the converting efficiencies versus frequency are calculated and plotted in Fig. 7(c), and the maximum converting efficiency can reach 82% at the central frequency 0.3 THz. The gray shadow in Fig. 7(c) represents the frequency band in which the converting efficiencies are over 50%. The half-power bandwidth of the bifunctional meta-coupler for the RCP incidence is 0.0285 THz. For the LCP incidence, the ratio of the power carried by SSP to that incident on the coupling metasurface, is considered to evaluate the converting efficiency [29]. The integral of poynting vector is used to calculate the power carried by SSP and the "l" in Fig. 7(a) represents the integral position. The relations between the converting efficiencies and frequency are plotted in Fig. 7(c) and marked in blue. From the efficiency curve we can know that the meta-coupler maintains high efficiency in a relatively broad frequency band and the maximum converting efficiency can reach 70% at 0.3 THz. The frequency band with efficiency higher than 50% is marked with red shadow and the corresponding half-power bandwidth is 0.01 THz.

C. THE EFFECT OF GAUSSIAN BEAM WAIST-WIDTH

Having demonstrated the high efficiency of the designed meta-coupler, next we will study its working performance with different beam waist-widths which are defined as $w$. The inset in Fig. 7(d) illustrates the definition of waist-width. All the results in last section are calculated under the condition $w = 1.2 \lambda$. The converting efficiencies versus the beam waist-width $w$ are calculated at $f = 0.3$ THz, as illustrated by Fig. 7(d). In the range from $0.7 \lambda$ to $1.7\lambda$, the bifunctional meta-coupler can maintain relatively high efficiency (>48%) for both the LCP and RCP incidence. In this range, the efficiencies of anomalous reflection have negative correlation with $w$ while the relations are inverse for SSP excitation. This is because, for the anomalous reflection, the higher efficiency requires more EMWs impinging on the coupling metasurface. So a wider Gaussian beam always corresponds to a lower converting efficiency. However, for the SSP excitation, the higher efficiency requires more uniform incident electric field amplitude distribution on the coupling metasurface. So a relatively wider Gaussian beam always corresponds to a higher converting efficiency. When the waist-width $w$ is out of this range, there will be evident drops in converting efficiencies.

V. CONCLUSION

This paper proposes a CP-controlled bifunctional meta-coupler for independent SSP excitation and beam steering. The normally incident waves are converted into surface waves and anomalously reflected under the illumination of the LCP and RCP waves, respectively. The respective converting efficiency can reach 82% and 70% for the RCP and LCP incidence and their corresponding half-power bandwidth is 0.0285 and 0.01 THz. By elaborately adjusting the resonant phase and rotation angle of each unit cell, many other diversified functionalities can also be integrated into the SSP meta-coupler. Our work can provide another degree of freedom in polarization-controlled unidirectional excitation of SSP and a new way of encoding the polarization information in different electric field modes.

**REFERENCES**
[1] H. Chu, J. Qi, S. Xiao, and J. Qiu, "A thin wideband high-spatial-resolution focusing metasurface for near-field passive millimeter-wave imaging," *Appl. Phys. Lett.,* vol. 112, no. 17, p. 174101, Apr. 2018.
[2] A. Pors, M. G. Nielsen, R. L. Eriksen, and S. I. Bozhevolnyi, "Broadband focusing flat mirrors based on plasmonic gradient metasurfaces," *Nano lett.*, vol. 13, no. 2, pp. 829-834, Jan. 2013.
[3] X. Li, S. Xiao, B. Cai, Q. He, T. J. Cui, and L. Zhou, "Flat metasurfaces to focus electromagnetic waves in reflection geometry," *Opt. lett.*, vol. 37, no. 23, pp. 4940-4942, Dec. 2012.
[4] M. Khorasaninejad, W. T. Chen, R. C. Devlin, J. Oh, A. Y. Zhu, and F. Capasso, "Metalenses at visible wavelengths: Diffraction-limited focusing and subwavelength resolution imaging," *Science*, vol. 352, no. 6290, pp. 1190-1194, Jun. 2016.
[5] F.-Y. Han, T.-J. Huang, L.-Z. Yin, J.-Y. Liu, and P.-K. Liu, "Superfocusing plate of terahertz waves based on a gradient refractive index metasurface," *J. Appl. Phys.*, vol. 124, no. 20, p. 204902, Nov. 2018.
[6] J.-J. Liang, G.-L. Huang, J.-N. Zhao, Z.-J. Gao, and T. Yuan, "Wideband Phase-Gradient Metasurface Antenna With Focused Beams," *IEEE Access*, vol. 7, pp. 20767-20772, Feb. 2019.
[7] A. M. Wong, P. Christian, and G. V. Eleftheriades, "Binary Huygens' Metasurfaces: Experimental Demonstration of Simple and Efficient Near-Grazing Retroreflectors for TE and TM Polarizations," *IEEE Trans. Antennas Propag.*, vol. 66, no. 6, pp. 2892-2903, Mar. 2018.
[8] V. Popov, F. Boust, and S. N. Burokur, "Controlling diffraction patterns with metagratings," *Phys. Rev. Appl.*, vol. 10, no. 1, p. 011002, Jul. 2018.
[9] L.-Z. Yin, T.-J. Huang, F.-Y. Han, J.-Y. Liu, and P.-K. Liu, "Terahertz multichannel metasurfaces with sparse unit cells," *Opt. Lett.*, vol. 44, no. 7, pp. 1556-1559, Mar. 2019.
[10] V. Asadchy et al., "Flat engineered multichannel reflectors," *Phys. Rev. X*, vol. 7, no. 3, p. 031046, Sep. 2017.
[11] Z.-Y. Liu, Q.-J. Wang, L.-R. Yuan, and Y.-Y. Zhu, "A multi-functional plasmonic metasurface for anomalous reflection and optical rotation on the basis of anisotropic building blocks," *J. Phys. D, Appl. Phys.*, vol. 50, no. 24, p. 245103, May 2017.

[12] F. Ding, Z. Wang, S. He, V. M. Shalaev, and A. V. Kildishev, "Broadband high-efficiency half-wave plate: a supercell-based plasmonic metasurface approach," *ACS nano*, vol. 9, no. 4, pp. 4111-4119, Mar. 2015.

[13] A. Pors, M. G. Nielsen, and S. I. Bozhevolnyi, "Broadband plasmonic half-wave plates in reflection," *Opt. lett.*, vol. 38, no. 4, pp. 513-515, Feb. 2013.

[14] Q. Chen and H. Zhang, "Dual-Patch Polarization Conversion Metasurface-Based Wideband Circular Polarization Slot Antenna," *IEEE Access*, vol. 6, pp. 74772-74777, Nov. 2018.

[15] N. K. Grady et al., "Terahertz metamaterials for linear polarization conversion and anomalous refraction," *Science*, vol. 340, no. 6138, pp. 1304-1307, Jun. 2013.

[16] D. Wen et al., "Helicity multiplexed broadband metasurface holograms," *Nat. commun.*, vol. 6, p. 8241, Sep. 2015.

[17] L. Huang et al., "Three-dimensional optical holography using a plasmonic metasurface," *Nat. commun.*, vol. 4, p. 2808, Nov. 2013.

[18] X. Ni, A. V. Kildishev, and V. M. Shalaev, "Metasurface holograms for visible light," *Nat. commun.*, vol. 4, p. 2807, Nov. 2013.

[19] G. Zheng, H. Mühlenbernd, M. Kenney, G. Li, T. Zentgraf, and S. Zhang, "Metasurface holograms reaching 80% efficiency," *Nat. nanotechnol.*, vol. 10, no. 4, p. 308, Feb. 2015.

[20] K.-Y. Liu, W.-L. Guo, G.-M. Wang, H.-P. Li, and G. Liu, "A Novel Broadband Bi-Functional Metasurface for Vortex Generation and Simultaneous RCS Reduction," *IEEE Access*, vol. 6, pp. 63999-64007, Nov. 2018.

[21] F. Ding, A. Pors, Y. Chen, V. A. Zenin, and S. I. Bozhevolnyi, "Beam-size-invariant spectropolarimeters using gap-plasmon metasurfaces," *Acs Photonics*, vol. 4, no. 4, pp. 943-949, Feb. 2017.

[22] A. Pors, M. G. Nielsen, and S. I. Bozhevolnyi, "Plasmonic metagratings for simultaneous determination of Stokes parameters," *Optica*, vol. 2, no. 8, pp. 716-723, Aug. 2015.

[23] J. B. Mueller, K. Leosson, and F. Capasso, "Ultracompact metasurface in-line polarimeter," *Optica*, vol. 3, no. 1, pp. 42-47, Jan. 2016.

[24] S. Sun, Q. He, S. Xiao, Q. Xu, X. Li, and L. Zhou, "Gradient-index meta-surfaces as a bridge linking propagating waves and surface waves," *Nat. mater.*, vol. 11, no. 5, p. 426, Apr. 2012.

[25] W. Sun, Q. He, S. Sun, and L. Zhou, "High-efficiency surface plasmon meta-couplers: concept and microwave-regime realizations," *Light: Sci. Appl.,* vol. 5, no. 1, p. e16003, Jan. 2016.

[26] Y. Meng et al., "Dispersion engineering of metasurfaces for supporting both TM and TE spoof surface plasmon polariton," *J. Phys. D, Appl. Phys.*, vol. 51, no. 4, p. 045109, Jan. 2018.

[27] J. Wang et al., "High-efficiency spoof plasmon polariton coupler mediated by gradient metasurfaces," *Appl. Phys. Lett.*, vol. 101, no. 20, p. 201104, Nov. 2012.

[28] C. Wu, Y. Cheng, W. Wang, B. He, and R. Gong, "Ultra-thin and polarization-independent phase gradient metasurface for high-efficiency spoof surface-plasmon-polariton coupling," *Appl. Phys. Express*, vol. 8, no. 12, p. 122001, Nov. 2015.

[29] J. Duan et al., "High-efficiency chirality-modulated spoof surface plasmon meta-coupler," *Sci. rep.*, vol. 7, no. 1, p. 1354, May 2017.

[30] Y. Meng, H. Ma, M. Feng, J. Wang, Z. Li, and S. Qu, "Independent excitation of spoof surface plasmon polaritons for orthogonal linear polarized incidences," *Appl. Phys. A*, vol. 124, no. 10, p. 707, Sep. 2018.

[31] S. Liu et al., "Negative reflection and negative surface wave conversion from obliquely incident electromagnetic waves," *Light: Sci. Appl.*, vol. 7, no. 5, p. 18008, May 2018.

[32] T. Cai et al., "High - performance bifunctional metasurfaces in transmission and reflection geometries," *Adv. Opt. Mater.*, vol. 5, no. 2, p. 1600506, Nov. 2017.

[33] F. Ding, R. Deshpande, and S. I. Bozhevolnyi, "Bifunctional gap-plasmon metasurfaces for visible light: polarization-controlled unidirectional surface plasmon excitation and beam steering at normal incidence," *Light: Sci. Appl.*, vol. 7, no. 4, p. 17178, Apr. 2018.

[34] Ling, Y., Huang, L., Hong, W., Liu, T., Jing, L., Liu, W., and Wang, Z.,, "Polarization-switchable and wavelength-controllable multi-functional metasurface for focusing and surface-plasmon-polariton wave excitation", *Opt. Express*, vol. 25, no. 24, pp. 29812-29821, Nov. 2017.

[35] X. Yin, Z. Ye, J. Rho, Y. Wang, and X. Zhang, "Photonic spin Hall effect at metasurfaces," *Science*, vol. 339, no. 6126, pp. 1405-1407, Mar. 2013.

[36] W. Luo, S. Xiao, Q. He, S. Sun, and L. Zhou, "Photonic spin Hall effect with nearly 100% efficiency," *Adv. Opt. Mater.*, vol. 3, no. 8, pp. 1102-1108, Apr. 2015.

[37] Y. Yuan, K. Zhang, X. Ding, B. Ratni, S. N. Burokur, and Q. Wu, "Complementary transmissive ultra-thin meta-deflectors for broadband polarization-independent refractions in the microwave region," *Photonics Res.*, vol. 7, no. 1, pp. 80-88, Dec. 2019.

[38] H.-X. Xu et al., "Completely Spin-Decoupled Dual-Phase Hybrid Metasurfaces for Arbitrary Wavefront Control," *ACS Photonics*, vol. 6, no. 1, pp. 211-220, Jan. 2018.

[39] J. B. Mueller, N. A. Rubin, R. C. Devlin, B. Groever, and F. Capasso, "Metasurface polarization optics: independent phase control of arbitrary orthogonal states of polarization," *Phys. Rev. lett.*, vol. 118, no. 11, p. 113901, Mar. 2017.

[40] T.-J. Huang, J.-Y. Liu, L.-Z. Yin, F.-Y. Han, and P.-K. Liu, "Superfocusing of terahertz wave through spoof surface plasmons," *Opt. Express*, vol. 26, no. 18, pp. 22722-22732, Aug. 2018.

[41] Z. Han, Y. Zhang, and S. I. Bozhevolnyi, "Spoof surface plasmon-based stripe antennas with extreme field enhancement in the terahertz regime," *Opt. lett.*, vol. 40, no. 11, pp. 2533-2536, May 2015.